\documentclass[twocolumn,showpacs,preprintnumbers,amsmath,amssymb]{revtex4}
\usepackage{graphicx}
\usepackage{dcolumn}
\usepackage{bm}
\def\bea {\begin{eqnarray}}
\def\eea {\end{eqnarray}}
\def\ra {\rightarrow}
\def\be {\begin{equation}}
\def\ee {\end{equation}}

\begin{document}
\title
{Thermal radiation from an expanding viscous medium}
\author{Sukanya Mitra, Payal Mohanty, Sourav Sarkar and Jan-e Alam}
\medskip
\affiliation{Theoretical Physics Division, 
Variable Energy Cyclotron Centre, 1/AF, Bidhan Nagar,
Kolkata - 700064, India}
\begin{abstract}
The effects of viscosity on the space time evolution of  QGP
produced in nuclear collisions at RHIC energies have been studied. 
The entropy generated due to the viscous motion of the fluid 
has been taken into account in constraining the initial temperature by the
final multiplicity (measured at the freeze-out point).
The viscous effects on the photon spectra has been 
introduced  consistently through the evolution dynamics and
phase space factors of the particles participating in the
production process. We notice a stronger effect on the
photon spectra originating from QGP than hadronic matter. 
A detectable shift is observed in the space-time 
integrated $p_T$ distribution of photons due to 
dissipative effects.
\end{abstract}

\pacs{25.75.-q,25.75.Dw,24.85.+p}
\maketitle

\section{Introduction}
Nuclear collisions at Relativistic Heavy Ion Collider (RHIC) and
Large Hadron Collider (LHC) energies are aimed at creating a
thermalized state of quarks and gluons called quark gluon plasma (QGP).
The  weakly interacting picture of the QGP
stems from the perception of
asymptotic freedom of QCD at high temperatures and densities.
However, the experimental data  from RHIC~\cite{npa2005}, especially
the measured elliptic flow of hadrons
indicate that the matter produced in Au+Au collisions exhibit
properties which are more like a strongly interacting liquid than
a weakly interacting gas. 
The magnitude of the transport coefficients
can be used to understand the strength of the interaction within the QGP.
Therefore, the study of the transport properties of QGP and hot hadrons 
is of paramount importance in characterizing the matter formed in
heavy ion collisions (HIC) at relativistic energies.    
For example, the shear viscosity or the internal friction of the fluid
symbolizes the ability
to transfer momentum over a distance of about a mean free path.
Therefore, in a system where the constituents interact strongly
the transfer of momentum is performed easily
- resulting in lower values of $\eta$.
Consequently such a system may be characterized by a small value of
$\eta/s$ where $s$ is the entropy density.
On the other hand, for a weakly interacting system the momentum
transfer between the constituents become strenuous which gives rise
to large $\eta$.
The importance of viscosity also lies in the fact that
it damps out the variation in the velocity and makes the fluid flow
laminar. A very small viscosity (large Reynold number) may make the
flow turbulent.  A lower bound on the  value of $\eta/s$ has been
found using AdS/CFT~\cite{KSS} (see also ~\cite{sm}).

Collisions between nuclei at ultra-relativistic energies
produce charged particles - either in the hadronic or in the
partonic state, depending on the collision energy.
Interactions among these charged particles produce real photons.
Because of their nature of interaction, the
mean free path of photons in the medium (hadronic or partonic)
is large compared to the size of the system formed in HIC.
Therefore, photons emanating from such a system brings  out the 
information of the source point very efficiently~\cite{mclerran,gale,weldon}
(see~\cite{alam1,alam2,rapp} for review) and hence 
electromagnetic
probes (photons and lepton pairs) may play crucial role
in extracting the transport coefficients. 

The effects of viscosity on the photon spectra resulting from
HIC enter through two main factors: (i) the modification of 
the phase space factor due to the deviation of the system from equilibrium and
(ii) the space time evolution of the matter governed by dissipative
hydrodynamics.  One more  important issue deserves to be mentioned here.
Normally, the initial temperature ($T_i$) and the thermalization time 
($\tau_i$) are constrained by the  measured hadron multiplicity
($dN/dy$).
This approach is valid for a system where there is no viscous loss
and the time reversal symmetry is valid. However,
for a viscous system  the entropy at the freeze-out point (which is 
proportional to the multiplicity) contains the initially produced entropy as
well as the entropy produced during the space time evolution due to non-zero
shear and bulk viscosity. Therefore, the amount of entropy generated during 
the evolution has to be subtracted from the total entropy at the
freeze-out point and the remaining part which is produced initially should be 
used to estimate the initial temperature. Therefore, for a given 
$dN/dy$ (which is associated with the 
freeze-out point) and $\tau_i$ the magnitude of $T_i$ will
be lower in case of viscous dynamics compared to ideal flow.

Effects of viscosity on the transverse momentum
distribution of photons was earlier considered in~\cite{JPG,akc97}
and recently the interest in this field is 
renewed~\cite{akc,bhatt,dusling}. 
Beyond a certain threshold in collision energy the system is expected to be 
formed in QGP phase which will inevitably  make a
transition to  the hadronic matter later. The measured spectra contain
contributions from both QGP and hadronic phases.
Therefore, it becomes imperative to estimate
the photon emission with viscous effects from QGP as well 
as  hadrons and identify a kinematic window where 
photons from QGP dominate.  
While in some of the earlier works~\cite{akc,bhatt,dusling} 
contributions from hadrons were ignored, in others~\cite{JPG,akc97}
the effects of 
dissipation on the phase space factors were omitted. 
In the present work
we study the effects of viscosity on the thermal photon spectra
originating from QGP and hadronic matter and argue that photons
can be used as a very useful tool to estimate $\eta/s$ and hence
characterize the matter.

The paper is planned as follows. In the next section the kinetic
theory formalism for evaluating the photon emission rate in the independent
particle approximation~\cite{galekapusta} is discussed. 
In section III we  discuss the viscous effects on the phase space
distributions of the partons or hadrons participating in the photon 
production processes. Hydrodynamical evolution with viscous effects 
has been mentioned in section IV.  Section V is devoted to results
and section VI is dedicated to the summary  and discussion.
  
\section{Production of thermal photons}
The transverse momentum ($p_T$) distribution of photons from a 
reaction of the type: $1+2\rightarrow 3 +\gamma$  taking place in a thermal 
bath  at a  temperature, $T$ is given by~\cite{galekapusta}:
\begin{eqnarray}
E\frac{dR}{d^3p}=\frac{\mathcal{N}}{2(2\pi)^8}\int \frac{d^3p_1}{2E_1}
\frac{d^3p_2}{2E_2}\frac{d^3p_3}{2E_3}f_1 f_2 (1\pm f_3)\nonumber\\
\delta^{(4)}(p_1+p_2-p_3-p)\overline{\lvert\mathcal{M}\rvert ^2}
\label{eq1}
\end{eqnarray}
where $R$ is the rate of photon production per unit four-volume, 
$\cal{N}$ is the over all degeneracy for the reaction under consideration,
$p_i$, $E_i$  and $f_i(E_i)$ are the three momentum, energy  and 
thermal phase space factor of the particle $i$ (either parton or 
hadron). $\overline{|{\mathcal {M}}|^2}$ is the square of the invariant amplitude for the
process under consideration. After some straight forward algebra
 Eq.~\ref{eq1} can be simplified to (see Appendix A):
\begin{eqnarray}
\frac{dR}{d^2p_Tdy}=\frac{\mathcal{N}}{16(2\pi)^8}
\int p_{1T}dp_{1T}dp_{2T}d\phi_1dy_1dy_2\nonumber\\
f_1 f_2 (1\pm f_3)\times \frac{\overline{\arrowvert \mathcal{M}\arrowvert^2}}
{\arrowvert p_{1T}\sin(\phi_1-\phi_2)+
p_{T}\sin\phi_2 \arrowvert_{\phi_2=\phi_2^0}}
\label{eq2}
\end{eqnarray}

The collision of nuclei at RHIC and LHC energies is expected
to  produce
QGP. This state of matter, once created with high internal pressure 
will undergo rapid expansion and consequently will cool down to the
temperature, $T_c$  for QGP to hadron transition. In a first order phase 
transition scenario the system remains in a mixed phase of QGP and hadrons
at $T_c$ until a time where the entire QGP converts to hadrons. 
The thermal equilibrium may also be maintained in
the hot hadronic phase till the freeze-out  point  
(achieved at a temperature, $T_F$) 
where the mean free path of the hadrons 
is too large for collisions to take place. 

The  measured photon spectra
($dN/d^2p_Tdy$) is the yield obtained after 
performing the space time integration
over the entire evolution history - from the
initial state to the freeze-out point.  Therefore, Eq.~\ref{eq2}
needs to be integrated over the four volume to connect the
theoretical results with experiments:
\begin{equation}
\frac{dN}{d^2p_Tdy}\mid_{y=0}=
\sum_{i=Q,M,H}\int d^4x\left[\frac{dR}{d^2p_Tdy}\mid_{y=0}\right]_{i}
\label{eq3}
\end{equation}
where $i\equiv Q, M, H$ represents QGP, mixed (coexisting
phase of QGP and hadrons)
and hadronic phases respectively.
The effects of viscosity enter the photon spectra through the
space time evolution governed by the dissipative hydrodynamics and
the phase space factor, $f_i$'s in Eq.~\ref{eq2}.

\subsection{Thermal photons from QGP}
 The contribution from QGP
 to the spectrum of thermal photons
 due to annihilation ($q$$\bar{q}$$\rightarrow$$g$$\gamma$) and
 Compton ($q(\bar{q})g\rightarrow q(\bar{q})\gamma$)
 processes has been calculated in~\cite{kapusta,bair} using
 hard thermal loop (HTL) approximation~\cite{braaten}. Later, it was
 shown that photons from the processes~\cite{aurenche}:
 $g$$q$$\rightarrow$$g$$q$$\gamma$,
 $q$$q$$\rightarrow$$q$$q$$\gamma$,
 $q$$q$$\bar{q}$$\rightarrow$$q$$\gamma$
 and $g$$q$$\bar{q}$$\rightarrow$$g$$\gamma$
 contribute in the same order $O(\alpha\alpha_s)$ as
 the Compton and annihilation processes.
 The complete calculation of emission rate from QGP to order $\alpha_s$
 has been performed by resumming ladder diagrams in the effective
 theory~\cite{arnold}.
 However, in the present work we consider only the Compton and
annihilation processes for photon production.  We expect that
the shift in the photon spectra from the ideal to the viscous
scenario will not alter drastically with the replacement of
the Compton + annihilation rates by the rate obtained in Ref.~\cite{arnold}.

\subsection{Thermal photons from hadronic matter}
A set of  hadronic reactions with all possible
iso-spin combinations have been considered for the 
production of photons~\cite{we1,we2,we3,turbide}
from hadronic matter. 
The relevant reactions and decays for photon production are:
(i) $\pi\,\pi\,\ra\,\rho\,\gamma$, (ii) $\pi\,\rho\,
\ra\,\pi\gamma$ (with $\pi$, $\rho$, $\omega$, $\phi$ and $a_1$ in the
intermediate state~\cite{we3}), (iii)$\pi\,\pi\,\ra\,\eta\,\gamma$ and
(iv) $\pi\,\eta\,\ra\,\pi\,\gamma$,
$\rho\,\ra\,\pi\,\pi\,\gamma$ and $\omega\,\ra\,\pi\,\gamma$.
The corresponding vertices's are obtained
from various phenomenological Lagrangians described in detail
in Ref.~\cite{we1,we2,we3}.
The effect of hadronic dipole
form factors has been taken into account in the present
work as in~\cite{turbide}.
\section{Viscous correction to the distribution function}
We assume that the system is slightly away from equilibrium which relaxes back
to equilibrium through dissipative processes.
Here we briefly recall the main 
considerations leading to the commonly used form for the 
first viscous correction, $\delta$f to the phase space factor, $f$
defined as follows~\cite{teaney}:
\begin{eqnarray}
&&f(p)=f_0(1+\delta f)\nonumber\\
&=&f_0\left(1+\frac{p^{\alpha}p^{\beta}}{2T^3}\left[C \langle\nabla_{\alpha}u_{\beta}\rangle+
A\Delta_{\alpha\beta}\nabla.u\right]\right)
\label{disf1}
\end{eqnarray}
where $f_0$ is the equilibrium distribution function, 
$\langle\nabla_{\alpha}u_{\beta}\rangle\equiv\nabla_{\alpha}u_{\beta}+
\nabla_{\beta}u_{\alpha}-\frac{2}{3}\Delta_{\alpha\beta}\nabla_{\gamma}u^{\gamma}$,
$\Delta_{\alpha\beta}=g_{\alpha\beta}-u_{\alpha}u_{\beta}$,
$\nabla_{\alpha}=(g_{\alpha\beta}-u_{\alpha}u_{\beta})\partial^{\beta}$,
$u_{\mu}$ being the four-velocity of the fluid. The coefficients $C$ and $A$ can be 
determined in the following way. 
Substituting $f$ in the expression  
for stress-energy tensor $T^{\mu\nu}$ we get,
\begin{eqnarray}
T^{\mu\nu}&=&\int\frac{d^3p}{(2\pi)^3E}p^{\mu}p^{\nu}f_0(1+\delta f)\nonumber\\
&=&T_{0}^{\mu\nu}+\Delta T^{\mu\nu}
\label{eq5}
\end{eqnarray}
where $T^{\mu\nu}_0=(\epsilon +P)u^{\mu}u^{\nu}-g^{\mu\nu}P$ 
is the energy momentum tensor for ideal fluid.
From general considerations~\cite{Weinberg} the dissipative part can be
written as
\begin{equation}
 \Delta T^{\mu \nu}=\eta\langle\nabla^{\mu}u^{\nu}\rangle
+\zeta\Delta^{\mu\nu}\nabla\cdot u
\label{tmn_diss}
\end{equation}
Equating the part containing $\delta f$
from (\ref{disf1}) with (\ref{tmn_diss}), $C$ and $A$ can be expressed in terms of 
the coefficients of shear ($\eta$)
and bulk ($\zeta$) viscosity respectively
in terms of which the phase space distribution for the system 
can be written as: 
\be
f=f_{0}\left(1+\frac{\eta/s}{2T^3}p^{\alpha}p^{\beta}\langle\nabla_{\alpha}u_{\beta}\rangle-
\frac{\zeta/s}{5T^3}p^{\alpha}p^{\beta}\Delta_{\alpha\beta}\nabla\cdot u\right)
\label{disf2}
\ee

For a boost invariant expansion in (1+1) dimension this can be simplified to get,
\be
f=f_{0}[1+\delta f_{\eta}-\delta f_{\zeta}]
\label{eq7}
\ee
where
\be
\delta f_{\eta}=\frac{\eta/s}{3T^3\tau}(p_{T}^2-2p_z^{\prime \, 2})\nonumber\\
\label{eq8}
\ee
and
\be
\delta f_{\zeta}=\frac{\zeta/s}{5T^3\tau}(p_{T}^2+p_z^{\prime \, 2})\nonumber\\
\label{eq9}
\ee
where $p_z^\prime=m_T\sinh(y-\eta)$ is the $z$-component of the momentum in the fluid
co-moving frame.
The phase space distribution with viscous correction (\ref{eq7}) 
thus enters the production rate of photon 
through Eq.~\ref{eq2}.
\section{Expansion dynamics}
As mentioned before the $p_T$ distribution 
of thermal photons is obtained by integrating the emission rate
over the evolution history of the expanding fluid.  
Relativistic viscous hydrodynamics can be used as a tool
for the  space-time dynamics of the fluid. 



\begin{figure}[h]
\begin{center}
\includegraphics[scale=0.38]{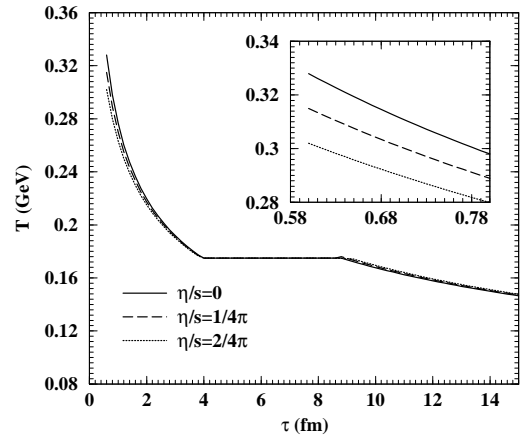}
\caption{Variation of temperature with proper time for 
different phases for various values of the shear viscosities. 
Inset shows the  effect of viscosity on the cooling of the
QGP phase (in an amplified scale) for different values of $\eta/s$.
}
\label{fig1}
\end{center}
\end{figure} 

For a (1+1) dimensional boost invariant expansion~\cite{bjorken} 
the evolution equation $\partial_\mu T^{\mu\nu}=0$,
can be written as~\cite{DG}:
\be
\frac{d\epsilon}{d\tau}+\frac{\epsilon+P}{\tau}=
\left(\frac{4}{3}\eta+\zeta\right)/\tau^{2}
\label{eq10}
\ee
where $P$ is the pressure and $\epsilon$ is the energy density.
We assume that the baryonic chemical potential 
is small in the central rapidity region for RHIC/LHC 
collision energies. Therefore,
the equation corresponding to the net baryon number 
conservation need not be considered in these situations.

We assume that the system achieves 
thermal equilibrium at a time $\tau_i$ after the collision at an
initial temperature $T_i$. 
With this initial condition 
and equation of state (EoS) $P=\epsilon/3$ 
the solution of Eq.~\ref{eq10} 
can be written as~\cite{JPG}: 
\begin{equation}
T=T_i\left(\frac{\tau_i}{\tau}\right)^{1/3}+
\frac{A_Q}{8 a_Q\tau_i}\left[\left(\frac{\tau_i}
{\tau}\right)^{1/3}-\frac{\tau_i}{\tau}\right]
\label{eq11}
\end{equation}
where $A_Q$=$\left(\frac{4}{3}\eta_{Q0}+\zeta_{Q0}\right)$, 
$\eta_{Q0}$=$\eta_Q/T^3$=$4a_Q(\eta/s)_Q$  and  
$\zeta_{Q0}$=$\zeta_Q/T^3$=$4a_Q(\zeta/s)_Q$.

Eq.~\ref{eq11} dictates the cooling of the QGP phase from its 
initial state   
to the transition temperature, $T_c$ at a time, $\tau_q$, when 
the QGP phase ends.
\par
In a first order phase transition scenario, the pure QGP phase is followed by 
a coexistence phase of QGP and  hadrons. 
The energy density, shear and bulk 
viscosities in the mixed phase can be written in terms of the corresponding
quantities of the quark and hadronic phases at temperature $T_c$ as follows
~\cite{JPG}:
\begin{eqnarray}
\epsilon_M(\tau)&=&f_Q\epsilon_Q(T_c)+(1-f_Q(\tau))\epsilon_H(T_c)\nonumber\\
\eta_M(\tau)&=&f_Q\eta_Q(T_c)+(1-f_Q(\tau))\eta_H(T_c)\nonumber\\
\zeta_M(\tau)&=&f_Q\zeta_Q(T_c)+(1-f_Q(\tau))\zeta_H(T_c)
\label{eq12} 
\end{eqnarray}
where $f_Q(\tau)$ ($f_H(\tau)$) indicates the
fraction of the quark (hadronic) matter in the mixed phase at a proper time
$\tau$.  We have
$\epsilon_Q(T_c)$=$3a_QT_c^4+B$, $\epsilon_H(T_c)$=$3a_HT_c^4$,
$a_Q=g_Q{\pi^2}/{90}$, $a_H=g_H{\pi^2}/{90}$,  $B$ 
is the bag constant,  $g_Q$ ($g_H$) denote statistical degeneracy 
for the QGP (hadronic) phase.
In the mixed phase the temperature remains constant but the 
energy density varies with time as the conversion of QGP to 
hadrons continues. This time variation is executed through $f_Q(\tau)$.
Substituting Eqs.~\ref{eq12} in Eq.~\ref{eq10} and solving for 
$f_Q(\tau)$  we get~\cite{JPG}, 
\begin{equation}
 f_Q=\frac{e^{-b/\tau}}{\tau}\int_{\tau_Q}^{\tau'}
\left[\frac{ce^{b/\tau'}}{\tau'}-ae^{b/\tau'}\right]d\tau'+
\frac{\tau_Q}{\tau}e^{(b/\tau_Q-b/\tau)}
\label{eq13}
\end{equation}
where a=$4\epsilon_H/(3\Delta\epsilon)$,
b=$[4(\eta_Q-\eta_H)/3+2(\zeta_Q-\zeta_H)]/\Delta\epsilon$,
c= $(\frac{4}{3}\eta_H+2\zeta_H)/\Delta\epsilon$ and
$\Delta\epsilon$=$\epsilon_Q-\epsilon_H$. Eq.~\ref{eq13} indicates
how the fraction of QGP in the co-existence phase evolves with time.

The variation of $T$ with $\tau$ in the hadronic phase can be 
obtained by solving Eq.~\ref{eq10} with the boundary condition $T=T_c$
and $\tau=\tau_H$, where $\tau_H$ is the (proper) time at which the mixed
phase ends {\it i.e.} when the conversion of QGP to hadronic matter 
is completed, 
\begin{equation}
T=T_c\left(\frac{\tau_H}{\tau}\right)^{1/3}+
\frac{A_H}{8 a_H\tau_H}\left[\left(\frac{\tau_H}
{\tau}\right)^{1/3}-\frac{\tau_H}{\tau}\right]
\label{eq14}
\end{equation}
Similar to QGP, $P=\epsilon/3$ has been used for hadronic  phase.
For a vanishing bulk viscosity ($\zeta=0$) the cooling of the
QGP is dictated by:
\begin{equation}
T=T_i\left(\frac{\tau_i}{\tau}\right)^{1/3}+
\frac{2}{3\tau_i}\left(\frac{\eta}{s}\right)_Q\left[\left(\frac{\tau_i}
{\tau}\right)^{1/3}-\frac{\tau_i}{\tau}\right]
\label{CLq}
\end{equation}
Similarly the time variation of temperature in the hadronic phase is given by:
\begin{equation}
T=T_c\left(\frac{\tau_H}{\tau}\right)^{1/3}+
\frac{2}{3\tau_H}\left(\frac{\eta}{s}\right)_H\left[\left(\frac{\tau_H}
{\tau}\right)^{1/3}-\frac{\tau_H}{\tau}\right]
\label{CLh}
\end{equation}
In a realistic scenario the value of $\eta/s$ may be different for
QGP~\cite{meyer,amy,wdow,das,greiner} and 
hadronic phases~\cite{itakura,dobado,ck,ktv}.
However, in the present work we take the same value of $\eta/s$ both for 
QGP and hadronic matter.



\section{Results}
In case of an ideal fluid, the conservation of entropy
implies that the rapidity density $dN/dy$ is a constant of motion
for the isentropic expansion \cite{bjorken}.
In such circumstances, the experimentally
observed (final) multiplicity, $dN/dy$ may be related to
a combination of the initial
temperature $T_i$ and the initial time $\tau_i$ as $T_i^3\tau_i$.
Assuming an appropriate value of $\tau_i$(taken to be $\sim 0.6 $ fm/c
in the present case), one can estimate $T_i$.

For dissipative systems, such an estimate is obviously inapplicable.
Generation of entropy during the evolution invalidates the role of
$dN/dy$ as a constant of motion. Moreover, the irreversibility
arising out of dissipative effects implies that estimation~ of the
~initial ~temperature from the final~ rapidity density~ is no longer
a trivial task. We can, nevertheless, relate the ~experimental
$dN/dy$ to the freeze-out temperature, $T_f$ and the freeze-out
time, $\tau_f$ by the relation,
\begin{equation}
\frac{dN}{dy}=\pi R_A^2 4 a_H T_f^3 \tau_f/\kappa
\label{eq17}
\end{equation}
where $R_A$ is the radius of the colliding nuclei(we consider $A A$
collision for simplicity) and $\kappa$ is a constant $\sim 3.6$ for
massless bosons.

To estimate the initial temperature for the dissipative fluid
we follow the following algorithm.
We treat $T_i$ as a parameter; for each $T_i$, we let the system evolve
forward in time under the condition of dissipative fluid dynamics 
(Eq.~\ref{eq10})
till a given freeze-out temperature $T_f$ is reached.
Thus $\tau_f$ is determined. We then compute $dN/dy$ at this instant of time
from eq.~\ref{eq17} and compare it with the experimental $dN/dy$. The
value of $T_i$ for which the calculated $dN/dy$ matches the
experimental number is taken to be the value of the  initial temperature.
Once $T_i$ is determined the evolution of the system  from the initial
to the freeze-out stage is determined by the Eqs.~\ref{eq11},~\ref{eq13}
and~\ref{eq14}.

In  Fig.~\ref{fig1} we display the variation of temperature with proper time. 
It is clear from the results shown in the inset (Fig.~\ref{fig1}) 
that initial temperature
for system which evolves with non-zero viscous effects is lower 
compared to the ideal case for a fixed $dN/dy$. 
Because for a non-viscous isentropic evolution scenario 
the multiplicity (measured at the freeze-out point) 
is fixed  by the initial entropy. 
However,  for a  viscous evolution scenario 
the generation  of entropy due to dissipative effects
contributes to the multiplicity. Therefore, for a given 
multiplicity (which is proportional
to the entropy) at the freeze-out point one requires lower 
initial entropy, hence initial temperature will be lower.    
It is also seen (Fig.~\ref{fig1}) that the cooling of the system is slower for
viscous dynamics because of the extra heat generated during the
evolution.
\subsection{Photon spectra}

In this section we present the shift in the $p_T$ distribution 
of the photons due to viscous effects.
The  integrand in Eq.~\ref{eq3} is a  Lorentz scalar,
consequently the Lorenz transformation of
the integrand from the laboratory to the co-moving frame of the
fluid can be effected by just transforming the argument, 
{\it i.e.} the energy of the photon ($E=p_T\cosh(y)$) in the laboratory 
frame should be replaced by $u_\mu p^\mu$ in the co-moving frame
of the fluid,  where $p^\mu$ is the four momentum  of the photon.

\begin{figure}[h]
\begin{center}
\includegraphics[scale=0.4]{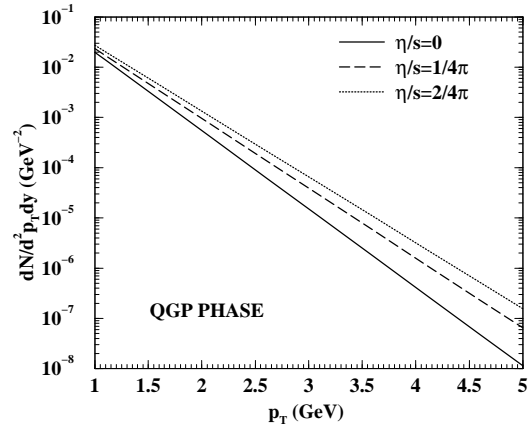}
\caption{Transverse momentum distribution of thermal photons from QGP for 
various values of $\eta/s$ in the scenario (i). 
}
\label{fig2}
\end{center}
\end{figure}

\begin{figure}[h]
\begin{center}
\includegraphics[scale=0.4]{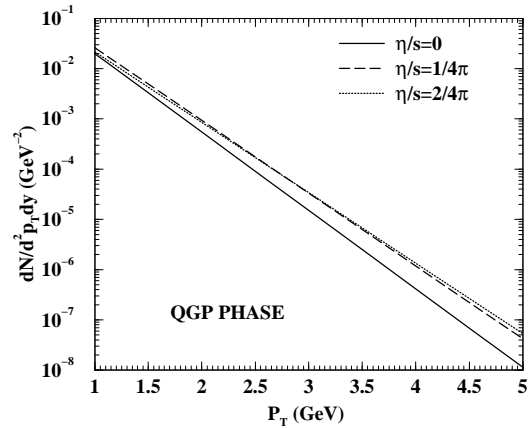}
\caption{Transverse momentum distribution of thermal photons from QGP 
for various values of $\eta/s$ in the scenario (ii). 
}
\label{fig3}
\end{center}
\end{figure}

The results presented here are obtained with vanishing bulk viscosity.
The effects of viscosity enters into the photon spectra 
through the phase space factor as well as through the space time
evolution. We would like to examine these two effects separately.
For convenience we define two scenarios: (i) where the effects
of viscosity on the phase space factor is included ($\delta f_\eta$ is non-zero
in Eq.~\ref{eq7}),  but the viscous effects on the evolution  are
neglected ($\eta=0$ in Eq.~\ref{eq10}) and scenario (ii) where
the effects of $\eta\neq 0$ are taken into account 
in the phase space factors as well as in the evolution dynamics.
The space time integrated photon yield originating from
the QGP in scenario (i) is displayed  in Fig.~\ref{fig2}. 
Note that the value of the initial temperatures for the
results displayed in Fig.~\ref{fig2} are same (for all $\eta/s$) because
the viscous effects on the evolution is ignored in scenario (i). 
The viscous effects on the $p_T$ distribution of 
the photons is distinctly visible. The higher values of  $\eta/s$ 
makes the spectra flatter through the $p_T$  dependence of the
correction, $\delta f_\eta$. 
Next we assess the effects of viscosity on photon spectra for scenario 
(ii).  In Fig.~\ref{fig3} we depict the photon spectra for various values of 
$\eta/s$. In this scenario the value of $T_i$ is lower for higher 
$\eta/s$ for reasons described above.  As a result the enhancement in 
the photon production due to change in phase space factor,
$\delta f_\eta$ is partially compensated by the  reduction in
$T_i$ for non-zero $\eta$, which is clearly seen in the 
results displayed in Figs.~\ref{fig2} and ~\ref{fig3}.

\begin{figure}[h]
\begin{center}
\includegraphics[scale=0.4]{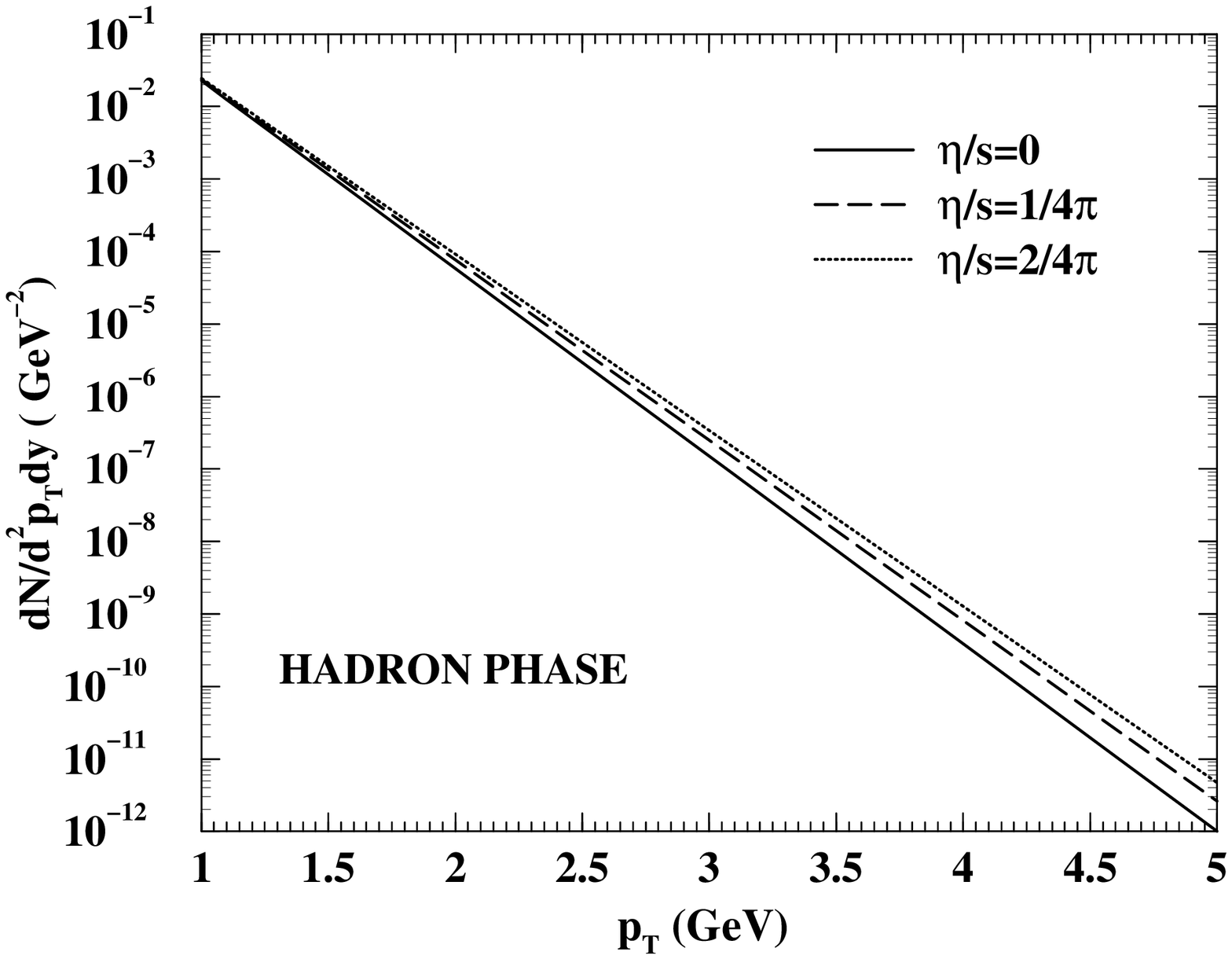}
\caption{Transverse momentum distribution of photons from thermal hadrons for 
various values of $\eta/s$ in the scenario (i). 
}
\label{fig4}
\end{center}
\end{figure}

\begin{figure}[h]
\begin{center}
\includegraphics[scale=0.4]{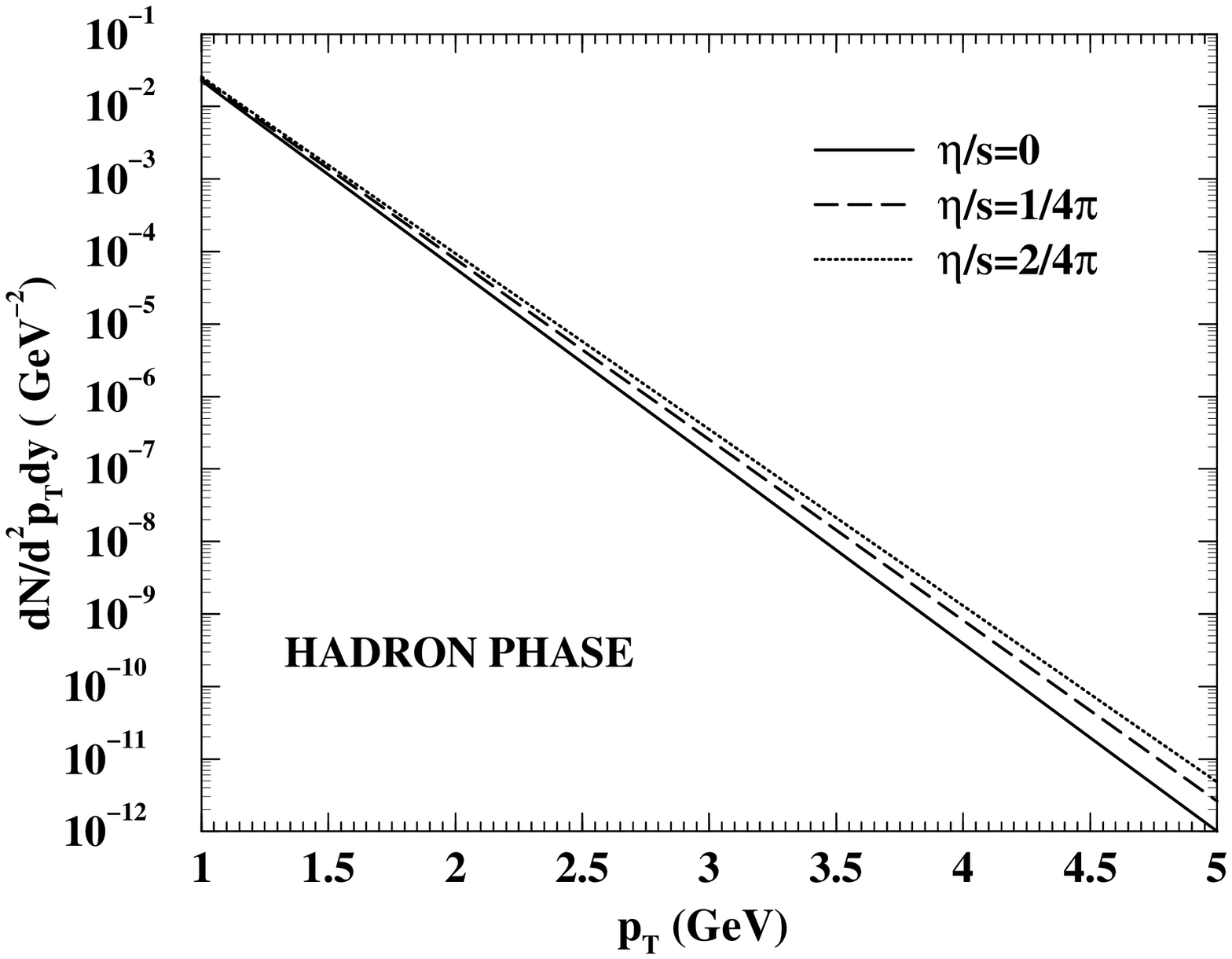}
\caption{Transverse momentum distribution of photons from thermal hadrons for 
various values of $\eta/s$ in the scenario (ii). 
}
\label{fig5}
\end{center}
\end{figure}

\begin{figure}[h]
\begin{center}
\includegraphics[scale=0.4]{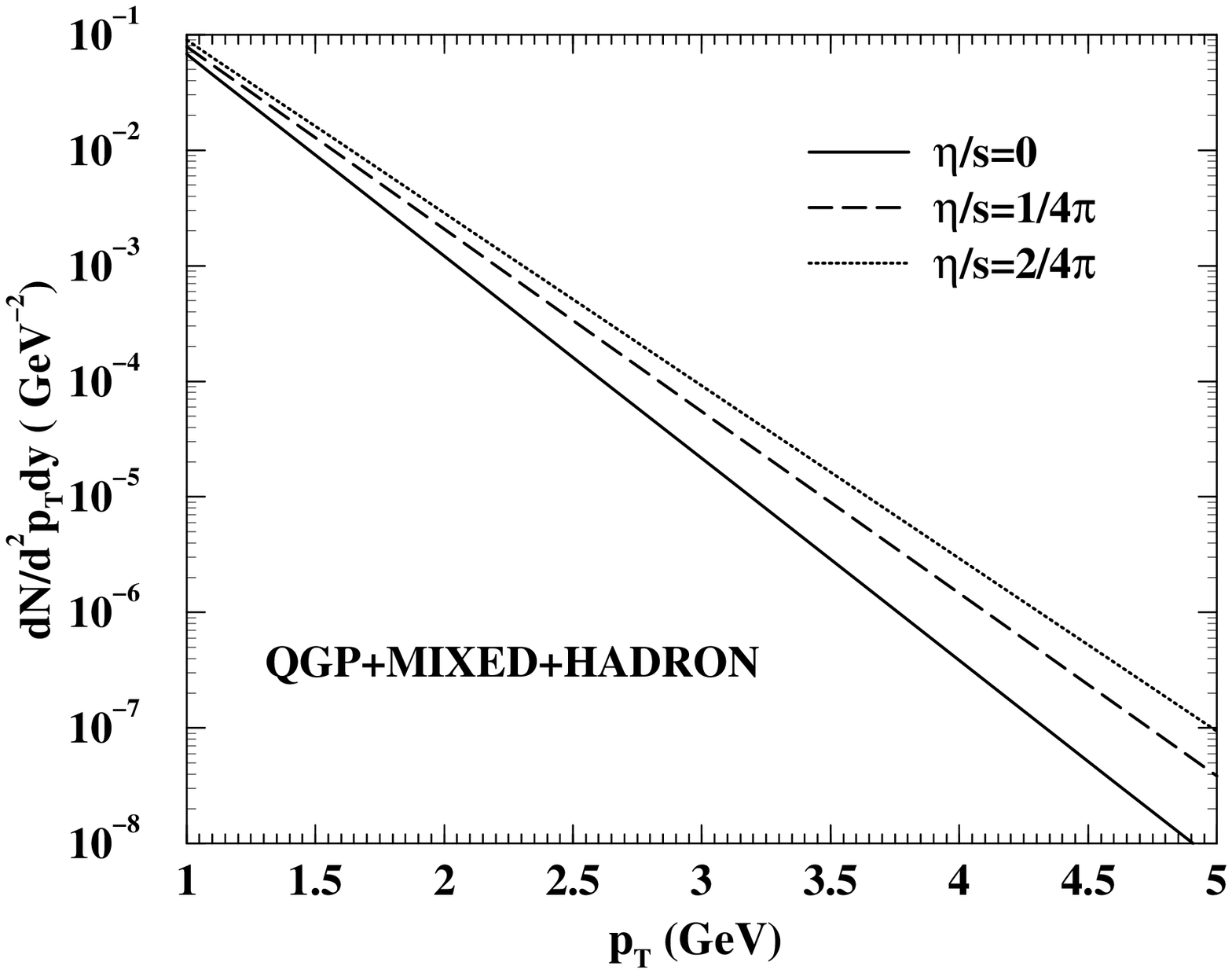}
\caption{Transverse momentum distribution of thermal photons from 
the entire evolution history of the system for 
various values of $\eta/s$ in the scenario (i). 
}
\label{fig6}
\end{center}
\end{figure}

\begin{figure}[h]
\begin{center}
\includegraphics[scale=0.4]{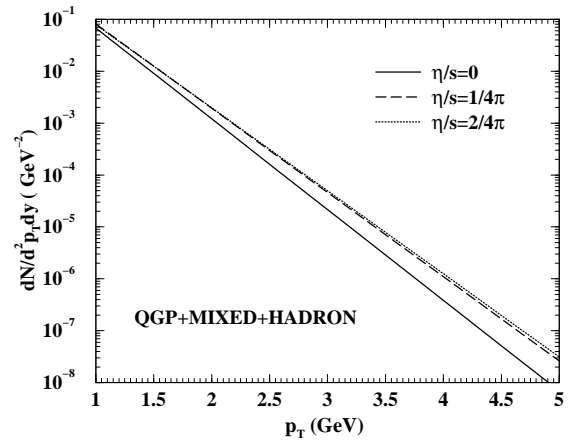}
\caption{Transverse momentum distribution of thermal photons from 
the entire evolution history of the system for 
various values of $\eta/s$ in the scenario (ii). 
}
\label{fig7}
\end{center}
\end{figure}

In Figs.~\ref{fig4} and ~\ref{fig5} we exhibit results for the 
hadronic phase for scenarios (i) and (ii) respectively. 
The effects of dissipation on the $p_T$ distribution of photons
from hadronic phase is qualitatively similar 
to the QGP phase
though the effect is stronger in the QGP phase
than in the hadronic phase. It is expected that the observed
shift in the photon spectra 
due to viscous effects may be detected in future
high precision experiments.
Finally in Figs.~\ref{fig6} and ~\ref{fig7} we plot the $p_T$ spectra 
of photons for the entire life time of the thermal system {\it i.e.}
the photon yield is obtained by summing up contributions from 
QGP, mixed and hadronic phases for different values of $\eta/s$.
The effect of viscosity for the scenario (i) is stronger than
than (ii).  



\section{Summary and Discussions}
We have studied the effects of viscosity on the evolving QGP
produced in nuclear collisions at RHIC energies. The generation of entropy
due to dissipation  on the final (experimentally measured)  
multiplicity has been taken into account. 
The initial temperature has been constrained
by the multiplicity (entropy) at the freeze-out point. 
The viscous effects on the photon spectra has been 
introduced  consistently through the evolution dynamics and
phase space factors of the particles participating in the
production processes.  It has been noticed that the effects of viscosity
on the photons from QGP is stronger than those originating from
hadrons. 
The QGP expected to be formed in Pb+Pb  collisions
at LHC energy will have longer life time and larger volume 
than that of RHIC {\it i.e.} the evolution dynamics 
at LHC will be dominated by QGP phase. Therefore, the chance
of estimating the value of $\eta/s$ from photon at LHC is brighter.
In view of the fact that the future experiments are progressing toward
precision measurement the shift in the 
$p_T$ distribution of photons due to dissipative effects
at LHC and RHIC may be detectable.

Before closing this section two comments are in order here. 
First, as mentioned before for the photon production rate  from
QGP we have used the Compton and annihilation processes. 
We have checked that the contribution from these 
two processes is down by a factor of 3-4 compared to the production
rate obtained from  the complete calculation of order $\alpha_s$ done in Ref.
~\cite{arnold}. Taking these higher order processes
into consideration in the present scenario involves a
reevaluation of the photon production rates with thermal
distribution factors containing viscous corrections.
Secondly, we have confined only to the longitudinal
flow of the  matter in the present work 
ignoring the transverse 
kick (blue shift) received by the photons from radial flow~\cite{assb}.
However, both 
these factors will affect the photon spectra from ideal as well
as dissipative scenarios in a similar fashion. Therefore,
we expect
the shift in the transverse momentum spectra  of thermal photons
in the presence of dissipative effects which
is the main focus of the present work, will be detectable even when a more
rigorous photon production rate along with transverse expansion is 
employed~\cite{inprogress}.

{\bf Acknowledgment:} P M and J A are supported by DAE-BRNS 
project Sanction No.  2005/21/5-BRNS/2455.

\section*{\textbf{Appendix A: Phase Space}} 
In this appendix we derive Eq. 2 from 1.The photon production rate from the process, $1+2\rightarrow3+\gamma$ is given by,
\begin{eqnarray}
E\frac{dR}{d^3p}=\frac{1}{2}\frac{\mathcal{N}}{(2\pi)^8}\int\frac{d^3p_1}{2E_1}\int\frac{d^3p_2}{2E_2}\int\frac{d^3p_3}{2E_1}\nonumber\\f_1(E_1)f_2(E_2)[1\pm
f_3(E_3)]\overline{|M|^2}\delta(p_1+p_2-p_3-p)
\end{eqnarray} 
Performing the $d^{3}p_{3}$ integration using the delta function and using $d^3p/E=p_Tdp_Tdyd\phi $ we get,

\begin{eqnarray}
E\frac{dR}{d^3p}=&&\frac{1}{16}\frac{\mathcal{N}}{(2\pi)^8}\int
p_{1T}dp_{1T}dy_1d\phi_1p_{2T}dp_{2T}dy_2d\phi_2\nonumber\\&&\frac{1}{E_3}
f_1(E_1)f_2(E_2)[1\pm f_3(E_3)]\nonumber\\&&\overline{|M|^2}\delta(E_1+E_2-E_3-E)
\label{eq19}
\end{eqnarray}
where $\phi_1$ and $\phi_2$ are the angles made by the transverse 
momenta of first and second particles with the transverse momentum of the emitted photon.
The momentum conservation along the $z$-direction: $p_{3z}=p_{1z}+p_{2z}-p_{z}$ can 
be written in terms of rapidity as:
\begin{eqnarray}
m_{3T}\sinh y_{3}=m_{1T}\sinh y_{1}+m_{2T}\sinh y_{2}-p_{T}\sinh y
\label{eq20}
\end{eqnarray} 
Now the energy, $E_3$ can be written as:
\be
E_{3}=m_{3T}\cosh y_{3}=\sqrt{m_{3T}^2+m_{3T}^2\sinh^2 y_3}
\label{eq21}
\ee
Substituting Eq.~\ref{eq20} in Eq.~\ref{eq21} we get,
\be
E_3=\sqrt{[(m_{1T}\sinh y_{1}+m_{2T}\sinh y_{2}-p_{T}\sinh y)^{2} 
+m_{3T}^2]}
\label{eq22}
\ee
Considering the energy conservation ($E_3=E_1+E_2-E$) and 
writing the energies in terms of rapidity ($E_i=m_{iT}\cosh y_i$) we get,
\be
E_3=m_{1T}\cosh y_1+m_{2T}\cosh y_2-p_T\cosh y
\label{eq23}
\ee
Equating Eqs.~\ref{eq22} and ~\ref{eq23} we have,
\begin{eqnarray}
m_{3T}=[m_{1T}^2+m_{2T}^{2}+p_{T}^{2}+2m_{1T}m_{2T}\cosh
(y_{1}-y_{2})\nonumber\\-2m_{1T}p_{T}\cosh (y_{1}-y)-2m_{2T}p_{T}\cosh (y_{2}-y)]^{\frac{1}{2}}
\label{eq24}
\end{eqnarray}

However, we also have,
\begin{eqnarray}
m_{3T}=& &(p_{3T}^2+m_3^2)^{\frac{1}{2}}\nonumber\\=&
&[(p_{1T}+p_{2T}-p_{T})^2+m_3^2]^{\frac{1}{2}}\nonumber\\=&
&[p_{1T}^{2}+p_{2T}^{2}+p_{T}^2+2p_{1T}p_{2T}\cos(\phi_{12})\nonumber\\&
&-2p_{T}p_{1T}\cos(\phi_{1})-2p_{T}p_{2T}\cos(\phi_{2})\nonumber\\& &+m_{3}^2]^{\frac{1}{2}}
\label{eq25}
\end{eqnarray}
where,
\begin{eqnarray} 
\cos(\phi_{12})= \cos(\phi_{1})\cos(\phi_{2})+\sin(\phi_{1})\sin(\phi_{2})
\end{eqnarray} 
Equating Eq.~\ref{eq24} with Eq.~\ref{eq25} leads to the expression,
\begin{eqnarray}
&
&[(p_{1T}\cos\phi_{1}-p_T)\cos\phi_2+p_{1T}\sin\phi_{1}\sin\phi_2]=\nonumber\\&
&\frac{1}{2p_{2T}}[(m_1^2+m_2^2-m_3^2)+2m_{1T}m_{2T}\cosh(y_1-y_2)\nonumber\\&
&-2m_{1T}p_{T}\cosh(y_1-y)-2m_{2T}p_{T}\cosh(y_2-y)\nonumber\\& &+2p_Tp_{1T}\cos\phi_1]
\label{eq26}
\end{eqnarray}
Solving Eq.~\ref{eq26} for $\phi_2$ one gets,
\begin{eqnarray}
\phi_2^0=\tan^{-1}(\frac{p_{1T}\sin\phi_1}{p_{1T}\cos\phi_1-p_T})-\cos^{-1}\frac{H}{2Rp_{2T}}
\end{eqnarray}
where,
\begin{eqnarray}
R=\sqrt{p_{1T}^2+p_T^2-2p_{1T}p_T\cos\phi_1}
\end{eqnarray}
and,
\begin{eqnarray}
H=& &(m_1^2+m_2^2-m_3^2)+2m_{1T}m_{2T}\cosh(y_1-y_2)\nonumber\\&
&-2m_{1T}p_{T}\cosh(y_1-y)-2m_{2T}p_{T}\cosh(y_2-y)\nonumber\\& &+2p_Tp_{1T}\cos\phi_1
\end{eqnarray}
Now we express  the argument of the delta function in Eq.~\ref{eq19} as function of $\phi_2$ as
\begin{eqnarray}
f(\phi_2)=&&E_1+E_2-E_3-E\nonumber\\=&&m_{1T}\cosh y_1+m_{2T}\cosh y_2-p_T\cosh y\nonumber\\&&-[m_{3T}^2+(m_{1T}\sinh y_1+m_{2T}\sinh y_2\nonumber\\&&-p_T\sinh y)^2]^{\frac{1}{2}}
\end{eqnarray}
and performing the $\phi_2$ integration in Eq.~\ref{eq19} we get,
\begin{eqnarray}
E\frac{dR}{d^3p}=&&\frac{1}{16}\frac{\mathcal{N}}{(2\pi)^8}\int^{\infty}_{0}
p_{1T}dp_{1T}\int^{\infty}_{0}dp_{2T}\nonumber\\&&\int^{\infty}_{-\infty}dy_1\int^{\infty}_{-\infty}dy_2\int^{2\pi}_{0}d\phi_1\nonumber\\&&
f_1(E_1)f_2(E_2)[1\pm f_3(E_3)]\nonumber\\&&\frac{\overline{|M|^2}}{|p_{1T}\sin (\phi_1-\phi_2)+p_T\sin \phi_{2}|_{\phi_{2}^{0}}}
\end{eqnarray}
with the constraint 
\begin{eqnarray}
|\frac{H}{2Rp_{2T}}|\leq 1
\end{eqnarray}
originating from $\mid \cos(\phi)\mid\leq 1$.

\end{document}